# Methods for Accurate Free Flight Measurement of Drag Coefficients

Elya Courtney, Amy Courtney, and Michael Courtney

BTG Research, 9574 Simon Lebleu Road, Lake Charles, LA, 70607
Michael_Courtney@alum.mit.edu

**Abstract:** This paper describes experimental methods for free flight measurement of drag coefficients to an accuracy of approximately 1%. There are two main methods of determining free flight drag coefficients, or equivalent ballistic coefficients: 1) measuring near and far velocities over a known distance and 2) measuring a near velocity and time of flight over a known distance. Atmospheric conditions must also be known and nearly constant over the flight path. A number of tradeoffs are important when designing experiments to accurately determine drag coefficients. The flight distance must be large enough so that the projectile's loss of velocity is significant compared with its initial velocity and much larger than the uncertainty in the near and/or far velocity measurements. On the other hand, since drag coefficients and ballistic coefficients both depend on velocity, the change in velocity over the flight path should be small enough that the average drag coefficient over the path (which is what is really determined) is a reasonable approximation to the value of drag coefficient at the near and far velocity. This paper considers these tradeoffs as well as practical considerations for obtaining accurate near and far velocity measurements and the impact of different sources of error (velocity, distance, time, atmospheric conditions, etc.) on the resulting accuracy of drag coefficients and ballistic coefficients. For a given level of accuracy of various quantities, the method of using near and far velocities usually produces drag coefficients with about half the uncertainty of the method using a near velocity and time of flight.

**Introduction**

Drag coefficients are of academic interest in both undergraduate laboratories and of practical interest in many areas of aerodynamics and applied science. In the field of ballistics, drag coefficients and/or the equivalent ballistic coefficients are commonly used to predict trajectories, wind drift, and kinetic energy retained downrange.

Theoretical values for drag and ballistic coefficients for specific projectiles provided by the manufacturers are often inaccurate (Courtney and Courtney, 2007; Litz, 2009; Halloran et al., 2012; Bohnenkamp et al., 2012). Furthermore, ballistic coefficients can depend on the bore of the rifle from which a bullet is shot (Bohnenkamp et al., 2011) and drag coefficients can depend on the gyroscopic stability of a projectile (Courtney and Miller, 2012a, 2012b; McDonald and Algren, 2003). If there is a practical or pedagogical reason to determine drag coefficients to an accuracy of better than 5-10%, then an accurate measurement technique is needed.

Two basic methods are commonly employed for determining drag coefficients of projectiles in free flight. The first method uses chronographs to measure a near and far velocity across a known distance and then uses a ballistic calculator (for example www.jbmballistics.com) and the environmental conditions to compute the ballistic coefficient from the velocities. Even if all the other values are measured perfectly, a 0.5% error in the near chronograph reading will result in a 3% error in the BC determination for a chronograph separation of 200 yards, standard atmospheric conditions, a G1 BC close to 0.400 and a muzzle velocity of 2800 fps. If the chronographs are separated by 100 yards, a 0.5% error in the near chronograph reading will result in a 6% error in the BC determination. Because BCs are inferred from relatively small velocity losses over distance, they are sensitive to errors in the velocity measurements. Some hobbyists also infer BCs from matching observed drops over a distance with output from a ballistic calculator. This method is notoriously inaccurate, because it is more sensitive to errors in many quantities than the more reliable methods discussed here.

The near velocity and time of flight technique is even more sensitive to errors in the velocity measurement. For example, even if all the other values are determined perfectly, a 0.5% error in the velocity determined by the near chronograph will result in a 6% error in the BC determination using the time of flight determined over 200 yards (same conditions as above) and a 12% error in the BC determination using a time of flight determined over 100 yards. Furthermore, a timing error of 0.001 s (about 0.5%) produces a BC error of 5%. The sensitivity of the time of flight technique to near velocity and timing errors is the main motivation for

# Methods for Accurate Free Flight Measurement of Drag Coefficients

preferring the method of measuring both near and far velocities for the accurate determination of ballistic and drag coefficients.

**Measurement Methods**
As with all experimental methods, there are tradeoffs between measurement accuracy, ease of use, and cost. Because of their widespread use by hobbyists, inexpensive optical chronographs are widely available, inexpensive, and some are designed well enough for laboratory use if due care is exercised to minimize sources of error and uncertainty.

*Lighting*
One of the largest sources of uncertainty in these hobbyist chronographs is related to the lighting. Optical chronographs are basically timers that start and stop the timer when each of two photodiodes detect the shadow of a bullet passing over a lens against the background of a light diffuser known as a skyscreen. The photodiode separation divided by the time to pass over them is the measured projectile velocity. When used outside, these skyscreens operate most uniformly either in cloudy conditions or when the sun is directly overhead so that there is one main shadow as the bullet passes directly between the lens and the skyscreen (in the vertical plane defined by the photodiode at the bottom and the skyscreen on top). Angles of the sun allowing sunlight to hit the lens directly can be a significant source of measurement error, as are other bright light sources from angles that can strike the lens or be the source of other shadows from passing projectiles.

After making efforts to reliably use these chronographs for several years, we switched to CED Millenium M2 chronographs with the optional infrared LED skyscreens. These may be used indoors and at night and for optimal accuracy, they should be shaded from the sun when used outside on sunny days. For around $300 for the complete chronograph (including LED skyscreens), this equipment has provided a cost effective approach to accurately measuring near and far velocities for free flight determination of drag coefficients.

*Chronograph Separation*
The chronograph separation should be chosen so that the projectile loses between 5% and 20% of its velocity over the separation distance. Velocity losses much below 5% make the resulting drag coefficients overly sensitive to velocity errors in the chronograph readings. Velocity losses above 20% make the drag coefficient determinations effectively averaged over a velocity range that is too large. Our experience tells us that velocity losses close to 10% represent a good tradeoff between these competing interests. This can be achieved with a chronograph separation of close to 100 yards for most bullets with a G1 BC close to 0.400, a muzzle velocity close to 2800 ft/s, and standard atmospheric conditions. Very high BC bullets (G1 BC = 0.700) shot in thin atmospheres may require closer to a 200 yard chronograph separation for the velocity loss to approach 10%. Faster bullets with lower BCs shot closer to sea level can lose enough velocity that good results may be obtainable with a 50 yard chronograph separation. Projectiles with higher drag (bbs, air rifle pellets, etc.) can lose 5-10% of their velocity over a chronograph separation of 10 yards.

As discussed below, it is also important to ensure that the projectiles pass through a relatively small window (a 1 inch to 2 inch diameter circle) near the center of the triangular area formed by the photodiode on bottom and the skyscreens on top so that the chronographs provide optimal accuracy. If the projectile launching system limits accuracy, it is better to choose a chronograph separation where the velocity loss is closer to 5% and the projectiles can remain well centered in triangular area of the far chronograph than to separate the chronographs the additional distance (which may be two to four times greater) required to achieve a 10-20% velocity loss. Once the chronographs are properly calibrated (see below) to remove systematic variations in velocity measurements, the uncertainty in resulting drag determinations can be reduced by increasing the sample size (number of shots) if needed.

It is preferred to use a chronograph spacing that yields closer to a 5% velocity loss if so doing facilitates shielding the projectile flight path from wind effects. Many of the combinations of projectile, velocity, and conditions under which we've tested bbs and pellets only have a 5% velocity loss over a 30 ft chronograph separation that is available indoors, but these projectiles are so light that this provides more accurate drag determinations than 60-100 ft chronograph separations provide outdoors if there are headwinds (that affect velocity and drag) or crosswinds (that move pellets in the target window).

Chronograph separation should be measured with a tape measure, because due care can yield separation uncertainties of a few inches over 100

# Methods for Accurate Free Flight Measurement of Drag Coefficients

yards. This is much less than the errors introduced by using laser range finders to measure chronograph separations. A 1 yard uncertainty in chronograph separation over 100 yards converts directly to a systematic 1% error in drag measurements, essentially using up the entire error budget with a single source of error if a 1% uncertainty is desired.

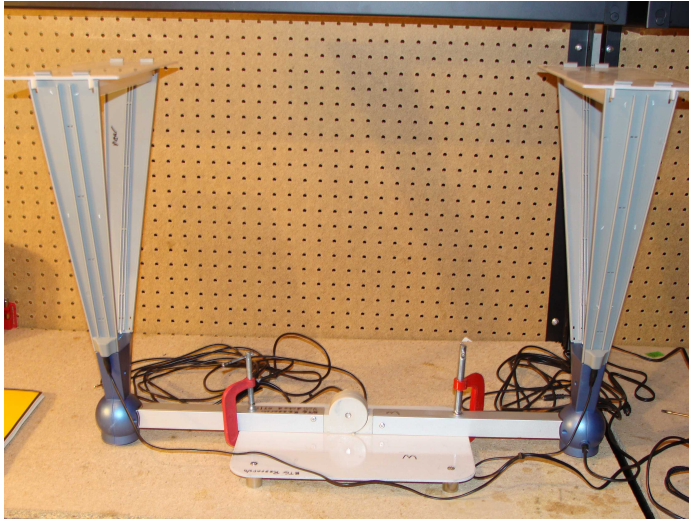

*Figure 1: CED Millenium M2 optical chronograph used for velocity measurements. Note addition of mechanical rigidity by clamping chronograph rail to steel plate to reduce relative motion of skyscreens and ensure they remain parallel during measurements.*

*Chronograph Skyscreen Planes Parallel*
The aluminum rail to which the photodiode assemblies and skyscreens are attached is sufficiently precise to ensure a constant separation of 24" and constant angular orientation, except for the hinge in the middle that allows the rail to fold in half for storage and transportation. In the absence of additional mechanical rigidity, one skyscreen is free to rotate upward relative to the other so that the two skyscreen planes would not be parallel. To ensure the skyscreen planes remain parallel, we use two C clamps to affix each side of the hinged rail to a rigid metal plate, as shown in Figure 1.

If the skyscreen planes are not parallel the distance between the skyscreens is not constant. For example, if the hinge angle is less than 180°, projectile paths near the top of the triangles are significantly shorter than projectile paths closer to the optical sensor near the bottom of the triangles. This results in velocity measurements of projectiles passing through the top of the triangular areas being systematically higher than their actual velocities.

*Chronograph Angle*
Some care needs to be taken to ensure the path of the bullet is perpendicular to the planes defined by the optical sensors. If the chronograph is canted relative to the bullet path, a small error is introduced because the chronograph assumes a 24" distance between optical planes. If the projectile path makes an angle θ with the normal to the optical planes, then the path distance from the near to the far optical plane is shortened to 24" cos(θ). For an angular misalignment of 2°, the path is shortened to 23.985" so that the velocities would be systematically high by 0.06%. In practice, we visually align the chronograph orientation angle with the path of the bullet by ensuring the centers of the optical sensors are aligned (left-right and top-bottom) from the perspective of the projectile launcher. Once the optical sensors and skyscreens themselves are aligned with the projectile launcher, we ensure that the position of the projectile launcher does not vary by marking the locations for the bipod feet on the bench or shooting table.

*Projectile Window Through Skyscreen Planes*
Keeping the projectile paths localized in a small area provides the most consistent, repeatable, and accurate velocity measurements possible. This is not a problem for relatively accurate projectile launchers using optical or laser sights as long as the far chronograph is at a distance to ensure all the projectiles pass through a 2" diameter projectile window centered directly over the optical sensors and high enough not to risk projectile impacts with the sensors. The projectile path over the sensors is easy to confirm using a paper target a short distance beyond the far skyscreen, maintaining care that the chronographs remain aligned, and marking the shooting bench or table to ensure the launch point remains constant.

*Chronograph Calibration*
The manufacturer's accuracy specification for the CED Millenium M2 chronograph is 0.3% in its velocity readings, and this inherent accuracy has been confirmed both in our laboratory and by independent parties (Litz, 2014). As discussed above, the relative error in a BC determination will be about 10 times larger than the relative error in velocity measurements. Consequently, a 0.3% error in velocity will contribute to a 3% error in drag determination. Another issue of concern is that the 0.3% error in velocity measurements has a strong systematic component, meaning that the errors are not random with respect to the true value, but rather tend to be biased either slightly high or slightly low. This systematic error cannot be reduced by averaging results from larger sample sizes (10-20 shots). We believe that the systematic error comes from dimensional and angular imperfections in the optical sensors and their spacing.

However, once both near and far chronographs are reading within 0.3% of the true velocities, drag determinations are more sensitive to relative errors between the far and near chronographs than they are to the absolute errors in each. In other words, if BOTH near and far chronographs read exactly 0.3% high, the resulting error in drag determinations will be much smaller than if the near chronograph reads 0.2% high and the far chronograph reads 0.1% low. Therefore, achieving a 1% uncertainty in resulting drag determinations does not require chronographs that both read within 0.1% of the true value, but only that the relative error between the chronographs be reduced to 0.1% of each other. This fact provides an opportunity to reduce errors in BC determinations by calibrating near and far chronographs with respect to each other.

The near and far chronographs are calibrated relative to each other by placing them in line, with minimal separation, and shooting though them. Each reading of the second chronograph is adjusted upward appropriately for the small loss of velocity (< 5 ft/s) over the two to four foot distance between chronographs. Then the average velocity of ten shots can be compared to determine systematic variations in the readings between the chronographs. The readings of the second chronograph are then multiplied by a calibration factor between 0.997 and 1.003 to bring the readings of both chronographs into agreement. In this manner, the variations between chronographs can be reduced to < 0.1%. The chronographs are calibrated using a standard bullet with a ballistic coefficient that is well known and known to have minimal (< 1%) shot to shot variations.

After the chronographs are calibrated each day, the far chronograph is moved to its experimental location. The accuracy of the experimental set-up can be further validated by measuring the drag coefficient of the standard bullet to confirm that it is in agreement with its well known value. This gives greater confidence to drag coefficients determined for other projectiles, but it is not an essential step if a projectile with a well known drag coefficient is not available.

*Atmospheric Conditions*
Atmospheric conditions impact drag measurements, because the most widely used formula relating drag coefficients to the force of aerodynamic drag models the aerodynamic drag force as being proportional to air density. In principle, there might also be a dependence on Reynolds number, but over the range of air densities commonly encountered near the earth's surface, the aerodynamics of most projectiles can be adequately modeled using the approximation that the only impact of atmospheric variables is through the proportionality of the drag force to air density. The accuracy of the approximation neglecting Reynolds number has recently been confirmed to an accuracy of 1-2% for supersonic projectiles up to Mach 2.8 (Courtney et al., 2014), and the accuracy of neglecting Reynolds number is also supported for subsonic data in the data presented here. Therefore, for practical purposes, the accuracy of free flight drag coefficients depends on the accuracy of air density as determined by atmospheric measurements of barometric pressure, temperature, and relative humidity.

*Barometric Pressure*
Since air density is proportional to barometric pressure and the drag force is proportional to air density, the relative error in drag coefficient is approximately equal to the relative error in the barometric pressure measurement. The Kestrel 4500 weather meter used to determine the atmospheric conditions measures barometric pressure to an accuracy better than 0.1%.

*Temperature*
At a given pressure, air density is inversely proportional to absolute temperature. Near standard conditions, a 1° F error in temperature measurement will yield approximately a 0.2% error in air density that will contribute a comparable error to the drag coefficient. Consequently, failure to measure the temperature with reasonable accuracy and to work under conditions where temperature is reasonably constant over the flight path may contribute

# Methods for Accurate Free Flight Measurement of Drag Coefficients

significant uncertainties to the resulting drag determinations. For longer flight paths, it may be prudent to station several weather meters along the flight path to accurately determine the average temperature over the path and to quantify how much the atmospheric conditions are changing over the flight path.

*Humidity*
Since water vapor is less dense (molecular weight of 18) than nitrogen (molecular weight of 28) and oxygen (molecular weight of 32), the air density is slightly lower in more humid air. In practical terms, this effect is small. For example, a difference of 50% in relative humidity only changes the atmospheric density by 0.32%. Consequently, the uncertainty in relative humidity by using the Kestrel 4500 weather meter only contributes a small uncertainty to drag determinations.

*Wind*
The environmental factor most likely to produce inaccuracies in drag measurements is wind. One simple way to obtain a first order estimate of the effect of the parallel component of wind on drag measurements is to estimate the additional distance a bullet must travel relative to still air when shooting into the wind.

For example, if shooting into a headwind, the headwind will slow the projectile more than still air. Suppose the chronographs are separated by 100 yards. A bullet with a near velocity of 2800 ft/s and a G1 BC of 0.400 will have a flight time close to 0.112 s. A 10 mph head wind may not seem important, but that converts to 14.7 ft/s; therefore, the bullet will travel through an effective additional distance in air of 1.65 ft while in flight, contributing a systematic error of approximately 0.6% to the drag determinations.

Effects of cross winds on drag determinations are less pronounced and mainly impact the measurements in the dual chronograph method by making it more challenging for the shooter to keep all the projectiles within a small target window over the optical sensors of the far chronograph.

Since wind conditions are highly variable on windy days, they can make significant contributions to shot to shot drag determinations. The most accurate drag determinations will be made at times when winds are minimal or by conducting measurements under indoor or sheltered conditions.

*Sample size*
If measurement errors are random, the uncertainty in the mean drag coefficient can be reduced by increasing the sample size. The standard error of the mean (SEM) generally decreases as the square root of the sample size, so that increasing the sample size by a factor of 4 can decrease the uncertainty in the mean by a factor of two. If a sample size of five is dominated by random errors and yields a SEM of 2%, then it is expected that a sample size of 20 would reduce the SEM to close to 1%.

**Data Analysis Methods**
Once accurate near and far velocities are obtained by the above experimental methods, the analysis proceeds depending on whether the drag coefficient or the ballistic coefficient is desired. The method to determine drag coefficients is described fully by Courtney et al. (2014) and uses the implicit definition of drag coefficient together with the work-energy theorem to obtain the formula,

$$C_d = \frac{2F_d}{Av^2 \rho} = \frac{m(v_f^2 - v_i^2)}{dAv^2 \rho},$$

where $F_d$ is the drag force, $m$ is the projectile mass, $d$ is the distance between chronographs, $\rho$ is the air density, $v$ is the average velocity over the interval (which has been substituted for the fluid flow velocity relative to the surface), $A$ is the cross sectional area of the projectile, and $v_f$ and $v_i$ are the final and initial velocities over the interval, respectively. Since most projectiles of interest have a circular cross section, the cross sectional area is computed as the area of a circle of appropriate diameter. There are several on-line calculators available for computing air density from ambient temperature, air pressure and humidity. We have validated the accuracy of the one built into the JBM ballistic calculator and QuickTARGET, so these are the ones we use.

The drag coefficient is determined for each shot at a given initial velocity. Then the mean and standard error of the mean of the drag coefficients are computed for all the trials taken at that velocity. The standard deviation of the mean quantifies the shot to shot variations from various sources including measurement uncertainties in the near and far velocities, small variations in shape of each projectile, and possible shot to shot variations in small deviations from point forward flight for non-spherical pro-

# Methods for Accurate Free Flight Measurement of Drag Coefficients

jectiles.

In principle, it is not possible to ascribe different sources of variation to specific causes with a high degree of certainty. However, over the years, we have noticed that some projectiles tend to yield larger standard deviations than others in drag determinations. Since the measurement method itself is common, we tend to think that projectiles with larger standard deviations (10%) in their drag measurements have larger real shot to shot variations in drag; whereas, the projectiles with the smallest (1%) standard deviations in drag determinations have their variations dominated by variations in the near and far velocity determinations of each shot. A 3% standard deviation is typical. Since SEM is the standard deviation divided by the square root of the sample size, an SEM of 1% can be obtained in many cases with a sample size of 10 shots.

Ballistic coefficients for a given drag model are obtained using the JBM ballistic calculator (www.jbmballistics.com ) or QuickTARGET. We have validated that these ballistic calculators are accurate implementations of the modified point mass method both by comparing the output with other reliable ballistic calculators that use the same method as well as by comparing our results with experimental data.

**Examples**

Figure 2 shows a graph of drag coefficient (Cd) vs. velocity for a 0.177 steel bb as measured with the dual chronograph method over a chronograph separation of 30 feet in an indoor range. The error bars consistently show uncertainties near 1% that result from sample sizes of 10 shots at each velocity. As expected the drag coefficient trends slowly upward from near 0.45 to above 0.50 as the velocity is increased toward the speed of sound. The best fit trend line (quadratic) is within (or nearly within) the estimated error of the data points, suggesting that the SEM provides a reasonable estimate of the error. These measurements agree with other drag coefficient measurements of spheres of comparable velocity and Reynolds number (Bailey and Hiatt, 1972).

The measurements resulting in the above drag determinations were made in an indoor facility in Colorado. Due to the mountain elevation, the ambient pressure was 22.93 in Hg. The temperature was 58° F. The relative humidity was 35.6%. These atmospheric factors yielded an air density of 0.9378 kg/m$^3$ with an estimated uncertainty of approximately 0.1% obtained by propagating the component uncertainties associated with the Kestrel 4500 weather meter. The standard deviations in the mean velocities from the 10 shots at each velocity were approximately 3%, resulting in SEMs just below 1% for the 10 shot sample sizes.

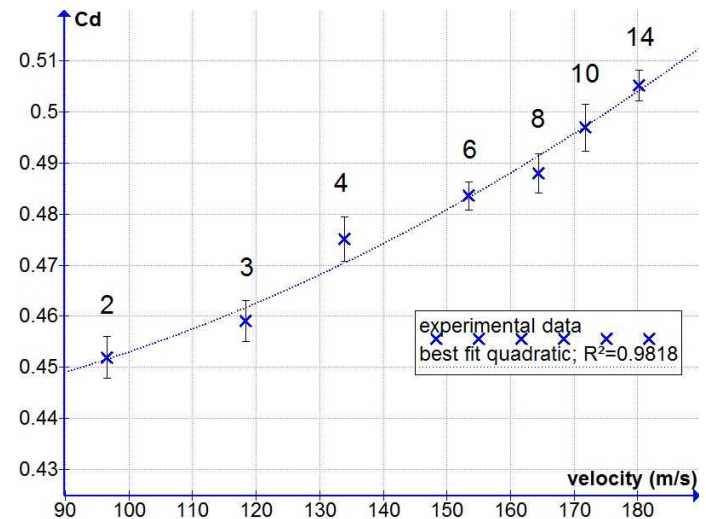

Fig 2: Drag coefficient vs. velocity for a 0.177 inch diameter steel bb (sphere) as determined by dual chronograph method over 30 feet. The number above each data point is the number of pumps used to generate the velocity in an inexpensive pump bb gun.

Figure 3 shows drag coefficients for different air rifle projectiles (pellets) fired from a spring pistol air rifle at two different air densities. Uncertainties vary from below 1% for the spheres and Crosman High Velocity Lead Free (CHVLF) pellets to above 5% for the Crosman Destroyer Ex (CDEx), even though the sample sizes were 10 for all projectiles. Six of the eight projectiles had lower uncertainties in the higher air density, which may be attributed to the greater velocity loss over the 30 ft chronograph separation. The larger uncertainties could probably be reduced to near 1% by increasing sample sizes by the appropriate amount.

# Methods for Accurate Free Flight Measurement of Drag Coefficients

It is notable that the drag values are within the error bars of each other for all projectiles. This supports both the idea that using the SEM as an error estimate is reasonable for the above method and that the drag coefficients may depend on velocity but do not depend strongly on Reynolds number for a given projectile near the surface of the earth.

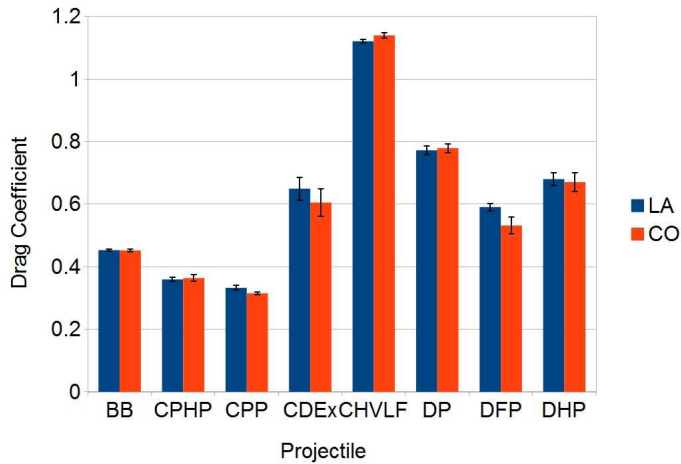

Fig 3: Drag coefficients for 8 different .177 inch diameter projectiles (bbs and pellets from an air rifle) as determined by dual chronograph method over a separation of 30 feet. Error bars show standard error from the mean for 10 shots. Drag coefficients are shown obtained with two different experiments, one at a high air density (1.115 kg/m$^3$) near sea level in Louisiana (LA) and the other at a low air density (0.9373 kg/m$^3$) at a mountain elevation in Colorado (CO).

We tend to think that projectiles with greater uncertainties either wobble more in flight or are not manufactured within the same dimensional tolerances as the projectiles with smaller uncertainties. The small uncertainties of the drag coefficients of the bbs supports this idea, because (being spherical) this is the only projectile for which the drag is independent of the orientation in flight. This is also supported by the observation that the Crosman Premier Hollow Point (CPHP) appears to have the same shape as the Daisy Hollow Point (DHP), but has a much smaller drag coefficient. If two projectiles have the same shape, well stabilized point forward flight with minimal wobble (pitch and yaw) would likely produce the same drag coefficients.

It is also notable that only the CPHP and the Crosman Pointed Premium have smaller drag coefficients than the steel sphere (bb). The high drag of the CHVLF is understandable because it has a flat tip nearly as wide as the projectile diameter.

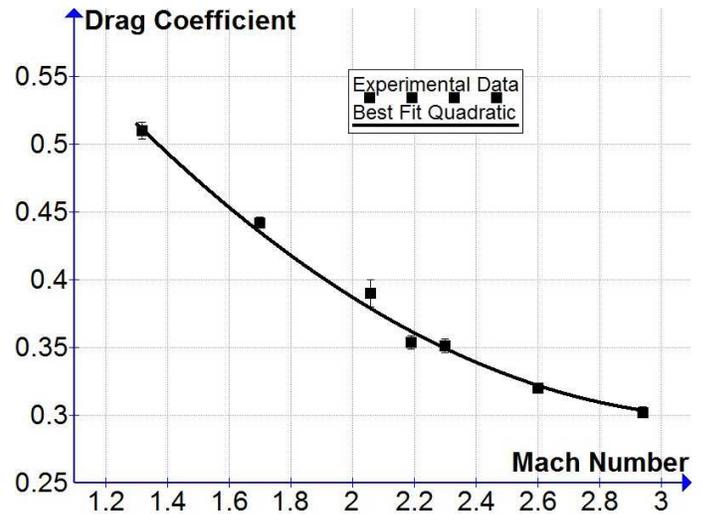

Fig 4: Drag coefficients for a supersonic .224" diameter 40 grain plastic tipped bullet over a range of Mach numbers. Error bars show standard error.

Figure 4 shows drag coefficients for a 40 grain plastic tipped bullet over a range of supersonic Mach numbers. Uncertainties tend to be 1-2% for 10 shots at each velocity. Over the years, we have noticed a trend that plastic tipped bullets tend to have lower shot to shot variations in their drag than other bullet designs. All projectiles have a trend where they experience a transonic drag rise as velocity is increased from subsonic to supersonic, and then the drag coefficient tends to decrease as the velocity is increased further above the speed of sound. Drag measurements for this bullet are consistent with the trend and the fact that a quadratic fit to the raw data passes through or very near to all the error bars suggests both that the SEM provides a reasonable error estimate and that the quadratic best fit curve is reasonable for interpolating drag coefficient values at Mach numbers in between actual drag measurements.

## Discussion

Most trajectory computations that account for air drag require knowing the drag coefficient as a function of velocity, Cd(v), and then integrating the equations of motion in a manner that accounts for the variation of drag coefficient with velocity. This requires either a look-up table or an interpolating function that returns a value of Cd for all the velocities across a range of

# Methods for Accurate Free Flight Measurement of Drag Coefficients

interest. In small arms ballistics and artillery, the requirement for a Cd value for a range of velocities is often fulfilled by the use of a drag model that specifies the shape of a curve describing what the function Cd vs. velocity looks like and a ballistic coefficient which serves as a scaling factor such that the drag coefficient is inversely proportional to the ballistic coefficient.

For a ballistic coefficient to be meaningful, one must specify the corresponding drag model. The G1 and G7 drag models are common in small arms ballistics today, and most ballistics calculators will compute a trajectory using the modified point mass model to numerically integrate the equations of motion given the initial conditions. Conversely, the near and far velocity information can be used in an appropriate ballistic calculator to compute the ballistic coefficient for a given drag model.

The method presented above is easily adaptable for academic use in introductory physics or engineering laboratories and provides a method for accurate measurement of drag coefficients within the meaning of the defining equation of drag coefficient. The logical reasoning should be clear to the introductory student. Near and far velocities are measured with accuracy close to 0.1% and then used to calculate near and far kinetic energies. The work-energy theorem is then used to compute an average drag force over the interval, and then a simple equation yields the drag coefficient from the defining equation. Several important principles of error propagation are apparent in the experimental method and features like a 1% uncertainty in drag coefficient requiring a 0.1% uncertainty in velocity are likely unexpected.

The above method is also adequate for accurate determination of drag and ballistic coefficients in small arms ballistics. BCs accurate to 1% are adequate for computing trajectories and retained energy downrange, as well as studying phenomena such as gyroscopic stability (Courtney and Miller, 2012a), pitch and yaw (Courtney et al., 2012), and dependence of BC (or Cd) on various experimental factors (Bohnenkamp et al., 2011).

Contemporary laboratories without budgetary restrictions often use RADAR methods to accurately determine drag coefficients over a projectile flight path, and there is no doubt that these expensive systems are easier to use and provide quick and accurate drag determinations over a range of velocities, since they can track projectiles downrange as velocity decays with distance. However, the method presented here can yield accurate drag coefficients for capital investments within the budgets of many educational laboratories and research facilities with more modest means (under $1000).

**Appendix**

Table 1 shows the G1 ballistic coefffficients measured for the 0.177" diameter bbs and air rifle pellets, along with the initial velocities at which they were determined. One simple way to interpret the meaning of G1 ballistic coefficients is as the fraction of 1000 yards over which the projectile will lose about 50% of its initial kinetic energy (or 71% of its initial velocity) in a standard atmosphere. For example, the Avanti bb, with a G1 BC of 0.01122 will lose 50% of its initial energy in about 11 yards, and the Crosman HV Lead Free pellet with a G1 BC of 0.00309 will lose 50% of its initial energy in only 3 yards.

| Projectile | G1 BC | Uncertainty | Vi (ft/s) |
|---|---|---|---|
| Avanti BB | 0.01122 | 0.00008 | 522 |
| Crosman Premier Hollow Point | 0.0221 | 0.0004 | 506 |
| Crosman Pointed Premium | 0.02083 | 0.0004 | 508 |
| Crosman Destroyer Ex | 0.01104 | 0.0006 | 543 |
| Crosman HV Lead Free | 0.00309 | 0.00001 | 789 |
| Daisy Pointed | 0.00859 | 0.0002 | 556 |
| Daisy Flat Point | 0.01276 | 0.0003 | 555 |
| Daisy Hollow Point | 0.01089 | 0.0003 | 482 |

*Table 1: G1 BCs, along with uncertainties (SEM) and initial velocities for the bbs and seven 0.177" diameter air rifle pellets measured with the method described in this paper.*

**Acknowledgements**

This work was funded by BTG Research (www.btgresearch.org) and the United States Air Force Academy. The authors are grateful to Colorado Rifle Club and Louisiana Shooters Unlimited where the experiments were performed. The authors appreciate valuable feedback from several reviewers which has been incorporated into revisions of the manuscript.

# Methods for Accurate Free Flight Measurement of Drag Coefficients